\documentclass[aps,prl,preprint,superscriptaddress,nofootinbib,floatfix,showkeys]{revtex4-1}

\usepackage{amsmath}
\usepackage{amssymb}
\usepackage{amsfonts}
\usepackage{graphicx}
\usepackage{xcolor}
\usepackage{xspace}
\usepackage{hyperref}
\usepackage{caption}
\usepackage{subcaption}

\usepackage{tikz}
\usetikzlibrary{decorations.pathmorphing}
\usetikzlibrary{decorations.markings}
\usetikzlibrary{positioning, shapes, arrows.meta}

\newcommand{\GeV}{\,\mathrm{GeV}}

\newcommand{\hvqdiscode}{\texttt{HVQDIS}\xspace}
\newcommand{\powheg}{\texttt{POWHEG}\xspace}
\newcommand{\powbox}{\texttt{POWHEG-BOX-V2}\xspace}

\newcommand{\lhapdf}{\texttt{LHAPDF}\xspace}

\begin{document}

\title{Heavy-quark pair-production in DIS at NLO QCD \\ matched to a parton shower}

\author{Santiago Castro}
\affiliation{Institute for Theoretical Physics,
  ELTE E\"otv\"os Lor\'and University,
  P\'azm\'any P\'eter 1/A, H--1117 Budapest, Hungary}

\author{Clara Del Pio}
\affiliation{Physics Department,
  Brookhaven National Laboratory, Upton, NY 11973, USA}

\author{Adam Kardos}
\affiliation{Department of Experimental Physics, Institute of Physics,
  Faculty of Science and Technology, University of Debrecen,
  4010 Debrecen, PO Box 105, Hungary}
\affiliation{Institute for Theoretical Physics,
  ELTE E\"otv\"os Lor\'and University,
  P\'azm\'any P\'eter 1/A, H--1117 Budapest, Hungary}

\author{Sven-Olaf Moch}
\affiliation{II.\ Institut f\"ur Theoretische Physik, Universit\"at Hamburg,
  Luruper Chaussee 149, D-22761 Hamburg, Germany}

\author{Aris Spourdalakis}
\affiliation{Institute for Theoretical Physics,
  ELTE E\"otv\"os Lor\'and University,
  P\'azm\'any P\'eter 1/A, H--1117 Budapest, Hungary}

\begin{abstract}
We present theoretical predictions for heavy-quark pair-production in deep-inelastic scattering (DIS) at next-to-leading order (NLO) in quantum chromodynamics (QCD), matched to a parton shower in the \powheg framework. 
We revisit the NLO heavy-quark pair-production cross section and implement a consistent matching to parton-shower evolution, with careful treatment of heavy-quark mass effects and the avoidance of double counting between fixed-order and parton-shower radiation. 
In addition, we compare the virtual NLO corrections available in the literature to one-loop amplitudes obtained through massification in the small-mass limit. 
This provides an independent validation of the virtual contributions. The study is presently restricted to the gluon-initiated channel, which dominates the kinematic region of interest at the HERA collider and remains important for the future Electron-Ion Collider.
%
\end{abstract}

\maketitle

\section{Introduction}

Heavy-flavor production in DIS is a key test of QCD and a sensitive probe of parton distribution functions (PDFs) in the proton, 
and measurements at the HERA collider~\cite{H1:2012xnw,H1:2018flt} have made heavy-quark DIS an essential part of precision QCD studies. 
In neutral-current DIS, charm- and bottom-quark pair production proceeds mainly via photon–gluon fusion, where a virtual photon scatters off a gluon in the nucleon. 
Moreover, the heavy-quark mass introduces an additional scale, making the process especially sensitive to perturbative QCD dynamics. 
The reduced cross section measured in experiments can be compared with theoretical predictions for inclusive heavy-quark DIS structure functions, which are known up to approximate next-to-next-to-leading order (NNLO) accuracy in perturbative QCD; see~\cite{Kawamura:2012cr,Ablinger:2025awb} and references therein.

DIS charm- and bottom-pair production, however, relies on identifying heavy-flavored mesons ($D$- or $B$-mesons) in the final state, and on phase-space cuts in the fiducial volume of the detector, requiring knowledge of the fully exclusive production process, including higher-order quantum corrections. 
Fully differential QCD corrections at next-to-leading order (NLO) have long been available~\cite{Riemersma:1994hv,Harris:1995tu,Harris:1997zq}. 
 The advent of the future Electron–Ion Collider (EIC) promises new high-precision measurements that will significantly improve our understanding of DIS heavy-flavor production and motivate more refined theoretical predictions.

As a first step toward improved theoretical accuracy for heavy-quark DIS, we present predictions for heavy-quark pair production at NLO in QCD, matched to a parton shower. The framework chosen here is the \powheg{}  method~\cite{Nason:2004rx,Frixione:2007vw,Nason:2012pr,Alioli:2010xd}, which is widely used at the Large Hadron Collider (LHC). 
Our work follows the \powheg{} implementation of charged-current heavy-quark DIS, which has been accomplished in~\cite{Buonocore:2024pdv}. The neutral- and charged-current DIS processes in the massless case have been studied in~\cite{Banfi:2023mhz}.
Here, we revisit the NLO heavy-quark pair-production cross section and implement consistent matching to parton-shower evolution. We also compare the virtual NLO corrections available in the literature (\hvqdiscode~\cite{Harris:1997zq}) with one-loop amplitudes obtained through massification~\cite{Mitov:2006xs,Czakon:2007ej,Czakon:2007wk} in the small-mass limit. The current implementation is limited to the photon-gluon-fusion channel, which dominates in the kinematic range relevant for HERA and the EIC, with extensions to all partonic channels underway. We expect that our software implementation and predictions will support future studies of the scientific requirements and detector concepts for the EIC~\cite{AbdulKhalek:2021gbh}. They could also assist a reanalysis of HERA data, with a focus on particle-level cross sections ($D$-meson production) and related studies of fragmentation.

In this letter, we briefly recall the theoretical framework and describe the NLO implementation, together with its validation against \hvqdiscode. We then compare our predictions with charm cross-section measurements from H1 and ZEUS at HERA, followed by corresponding results for bottom-quark production. Next, we present hadron-level NLO predictions matched to \powheg and discuss predictions in the EIC kinematic regime. We conclude with a summary of our main findings and an outlook.

\section{Theoretical framework}

In this section, we set the notation for the theoretical description of neutral-current heavy-quark production in DIS and provide details on its implementation in \powbox~\cite{Alioli:2010xd}.

\subsection{Heavy quark pair production in DIS}
Let us recall the leading order (LO) DIS kinematics. We consider the production of a pair of heavy quarks from an initial state massless parton, and the following scattering of one of these massive quarks off a massless lepton, via the exchange of a photon of virtuality $Q^2$
\begin{equation}
    e^-(p_1) + H(P) \to e^-(p_3) + Q(p_4) + \bar{Q}(p_5) \, .
\end{equation}
We take into account only the gluon-initiated channel, leaving the quark channel, which is formally a $\mathcal{O}(\alpha_s)$ correction with respect to the gluon one, for future work. We define the DIS variables as
\begin{equation}
    Q^2 = -q^2 = -(p_1-p_3)^2 \, , \quad x_{\rm Bj} = \frac{Q^2}{2 P \cdot q} \, , \quad y = \frac{P \cdot q}{P \cdot p_1} \, ,
\end{equation}
where $P$ is the hadron four-momentum, $p_1$ and $p_3$ are the initial- and final-state lepton momenta, respectively. We denote with $p_2=x P$ the momentum of the incoming gluon, which carries a longitudinal fraction $x$ of $P$.

The differential cross section in $x_{\rm Bj}$ and $Q^2$ can be written as
\begin{equation}
    \frac{d^2 \sigma}{d x_{\rm Bj} \,  dQ^2} = \frac{2 \pi \alpha^2(Q^2)}{x_{\rm Bj} Q^4}  \left[ \left( 1+(1-y)^2\right) F_2(x_{\rm Bj},Q^2,m^2) - y^2 F_L(x_{\rm Bj},Q^2,m^2)\right] \, ,
\end{equation}
where the structure functions $F_2(x_{\rm Bj},Q^2,m^2)$ and $F_L(x_{\rm Bj},Q^2,m^2)$, which depend on the heavy quark mass $m$, are defined as in~\cite{Riemersma:1994hv}, and $\alpha$ denotes the electromagnetic coupling.
The numerical results here will be presented in terms of the reduced cross section
\begin{align}
    \sigma_{\rm red}^{Q\bar{Q}} &= \frac{d^2 \sigma}{d x_{\rm Bj} \,  dQ^2} \cdot \frac{x_{\rm Bj} Q^4}{2 \pi \alpha^2(Q^2) \left( 1+(1-y)^2\right)} \nonumber \\
    & = F_2(x_{\rm Bj},Q^2,m^2) - \frac{y^2}{1+(1-y)^2} F_L(x_{\rm Bj},Q^2,m^2) \, .
\end{align}

At NLO in QCD, the process receives real and virtual corrections, see~\cite{Laenen:1992zk,Laenen:1992xs}. 
The virtual corrections require renormalization, with the strong coupling $\alpha_s$ renormalized in the $\overline{\rm MS}$ scheme and the heavy-quark mass in the on-shell scheme (pole mass). 
We have re-computed the Born and real-emission amplitudes with the spinor-helicity method~\cite{Campbell:2023fjg}.
For the virtual corrections, we follow two different strategies: we can either construct the azimuth-averaged virtual from the transverse and longitudinal projections from the literature~\cite{Harris:1995tu} that we checked against \texttt{GoSam}~\cite{GoSam:2014iqq} in the $y\to 1$ limit, or start from the massless amplitudes, that we re-calculated, and apply a massification procedure. Studying the effect of massification at NLO also provides useful insights on the accuracy of this approximation, that could be used for the calculation of the relevant heavy-quark pair production amplitudes at NNLO, where the exact result is currently unknown.

The massified virtual amplitude can be obtained from the massless one as 
\begin{align}
    \left|\mathcal{M}^{(1)}_{m}\right\rangle 
    =
    \left|\mathcal{M}^{(1)}_{0}\right\rangle 
    +
    \boldsymbol{I}_m 
    \left|\mathcal{M}^{(0)}_{m}\right\rangle 
    -
    \boldsymbol{I}_0 
    \left|\mathcal{M}^{(0)}_{0}\right\rangle 
    \,,
\end{align}
where $\left|\mathcal{M}^{(1)}_{m}\right\rangle$ and $\left|\mathcal{M}^{(1)}_{0}\right\rangle$ are the massified and massless one-loop matrix elements, respectively, $\left|\mathcal{M}^{(0)}_{m}\right\rangle$ is the tree-level Born matrix element calculated with
a massive quark pair, while $\left|\mathcal{M}^{(0)}_{0}\right\rangle$ is the one with a massless quark pair.
$\boldsymbol{I}_0$ corresponds to the universal operator that embodies the infrared (IR) singular structure in the massless case~\cite{Catani:1998bh} and contains logarithms of the form $\ln\left(\mu^2 / s_{ij}\right)$, while $\boldsymbol{I}_m$ is the corresponding operator for the massive case~\cite{Catani:2000pi} and carries both logarithms of the form $\ln\left(\mu^2 / s_{ij}\right)$ and $\ln\left(\mu^2/m^2\right)$. To achieve the proper replacement of IR logs both operators need to be expanded up to and including $\mathcal{O}(\varepsilon^{0})$ terms. Using the corresponding Born matrix elements is essential to achieve the complete cancellation of  IR poles.

\subsection{\powheg implementation}

We describe here some further details on the implementation of the fully differential event generator for the simulation
of neutral-current heavy-quark DIS at NLO in QCD, consistently matched to a parton shower (PS). 

Beside the implementation of the analytic expressions for the Born, virtual and real
matrix elements, the fixed-order NLO calculation requires a subtraction scheme
for the infrared divergences. We work in the \powheg  framework, which uses the
Frixione-Kunszt-Signer (FKS) subtraction scheme~\cite{Frixione:1995ms}. A crucial ingredient is the momentum
mapping, that enables us to write the real phase space in terms of the product of the so-called underlying Born phase space configuration times a set of radiation variables. As noted in Ref.~\cite{Banfi:2023mhz,Buonocore:2024pdv}, the default mapping of the \powheg code is not suitable for electron-hadron interactions, as it leads to  distortions in the distributions of the leptonic variables, that should be preserved if one considers  only QCD corrections. Furthermore, altering the lepton momentum can result in modifying the fiducial cross section when going from NLO accuracy to the NLO+PS matching to the parton shower program~\cite{Buonocore:2024pdv}.

Given these considerations, we implement the mapping presented in Ref.~\cite{Buonocore:2024pdv}, that preserves 
 the momenta of both the initial and final state lepton by adapting the results of Ref.~\cite{Catani:1996vz,Banfi:2023mhz} to the case of FKS subtractions and to a massive lepton and quark in the final state. We straightforwardly generalize the implementation of Ref.~\cite{Buonocore:2024pdv} to a massive quark pair in the final state by requiring the following variables to be preserved in the mapping
 \begin{equation}
     x_{\rm Bj}, \quad Q^2 \; ({\rm or \; equivalently} \; y), \quad m_{Q\bar{Q}} \, ,
 \end{equation}
 with $m_{Q\bar{Q}}$ the invariant mass of the heavy quark-pair system. Notice that, while in Ref.~\cite{Buonocore:2024pdv} the charm quark mass is a fixed parameter, in the present case $m_{Q\bar{Q}}$ is the dynamical scale of the massive quark-pair.

\section{Validation against \hvqdiscode}
%
\begin{figure}[h]
  \centering
  \begin{subfigure}[t]{0.48\columnwidth}
    \includegraphics[width=\linewidth]{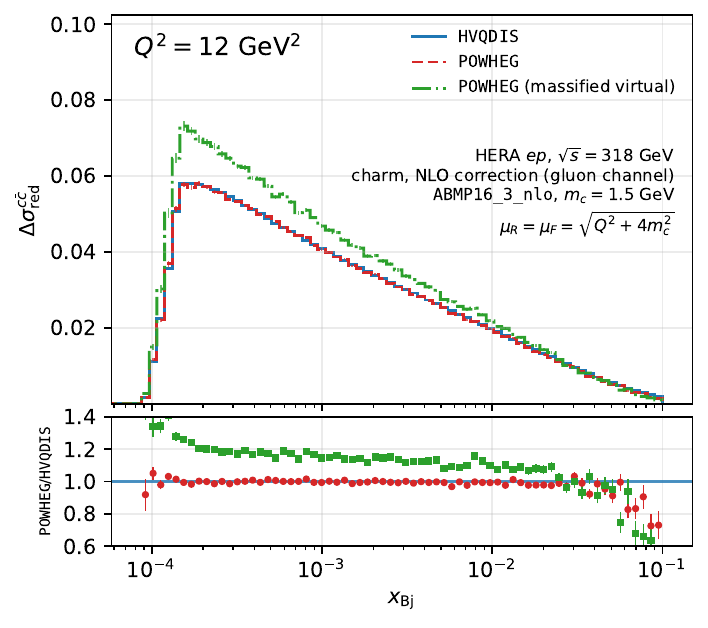}
    \caption{$Q^2 = 12\GeV^2$ (just above $4 m_c^2$).}
    \label{fig:nlo_hsmf_q2_12}
  \end{subfigure}\hfill
  \begin{subfigure}[t]{0.48\columnwidth}
    \includegraphics[width=\linewidth]{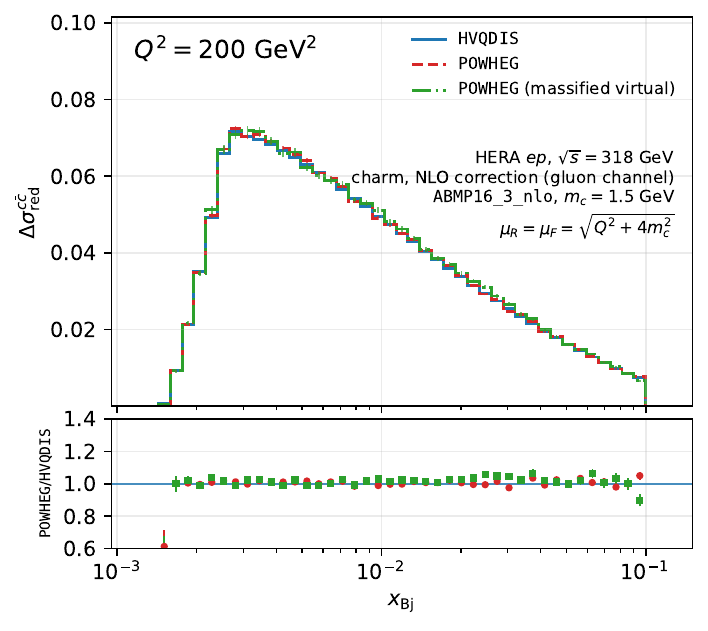}
    \caption{$Q^2 = 200\GeV^2$ ($m_c^2/Q^2 \approx 0.01$).}
    \label{fig:nlo_hsmf_q2_200}
  \end{subfigure}
  \caption{NLO correction to the reduced charm cross section
    $\Delta\sigma_{\rm red}^{c\bar{c}}$ as a function of $x_{\rm Bj}$ at two
    representative $Q^2$, from \hvqdiscode (solid), \powheg
    (dashed), and \powheg with a massified virtual amplitude
    (dash-dotted), with the two \powheg/\hvqdiscode ratios in the
    lower panels (see text).}
  \label{fig:nlo_hsmf}
\end{figure}

\begin{table}[t]
  \centering
  \begin{tabular}{lcccc}
\hline\hline
 & \multicolumn{2}{c}{Charm ($c\bar c$) [nb]} & \multicolumn{2}{c}{Bottom ($b\bar b$) [nb]} \\
Contribution & \texttt{HVQDIS} & \texttt{POWHEG} & \texttt{HVQDIS} & \texttt{POWHEG} \\
\hline
LO & $34.3539(1)$ & $34.3511(9)$ & $0.723844(2)$ & $0.723747(43)$ \\
NLO correction & $14.6432(6)$ & $14.6470(32)$ & $0.329347(4)$ & $0.329689(38)$ \\
LO\,+\,NLO & $48.9971(6)$ & $48.9980(33)$ & $1.053191(5)$ & $1.053436(57)$ \\
\hline\hline
\end{tabular}

  \caption{Total charm- and bottom-production cross sections (in nb)
    from \hvqdiscode and \powheg at LO and NLO. The two codes agree
    at the per-mille level on every entry for both flavors. The
    number in parentheses is the Monte-Carlo integration uncertainty
    on the last quoted digit(s).}
  \label{tab:xsec}
\end{table}

We validate our implementation against \hvqdiscode at the level of
both the total cross section (Table~\ref{tab:xsec}) and the
differential reduced cross section (Fig.~\ref{fig:nlo_hsmf}). Both
codes are evaluated in the gluon channel and we consider charm- and bottom-quark production. For the charm quark, the settings are $m_c = 1.5\GeV$,
\texttt{ABMP16\_3\_nlo}~\cite{Alekhin:2017kpj} (interfaced via
\lhapdf~\cite{Buckley:2014ana}),
$\mu_R=\mu_F=\sqrt{Q^2+4m_c^2}$, over
$2 \le Q^2 \le 2500\GeV^2$; the bottom quark validation uses the
corresponding $4$ light flavor  configuration ($n_f=4$, with $m_b = 4.5\GeV$,
\texttt{ABMP16\_4\_nlo},
$\mu_R=\mu_F=\sqrt{Q^2+4m_b^2}$). Throughout this work the
electromagnetic coupling is fixed at $1/\alpha = 137.04$,
following the original Harris--Smith convention~\cite{Harris:1995tu}.
Quoted uncertainties are Monte-Carlo
integration errors. For simplicity, the LO comparison against
\hvqdiscode uses the same NLO PDF set rather than a dedicated LO fit,
since the validation tests the matrix-element implementation rather
than the PDF.

In the charm case, the \powheg--\hvqdiscode
comparison agrees at the per-mille level throughout, validating the
matrix-element implementation. As an additional check, we also
compared the total cross section against the
results of Riemersma~et~al.~\cite{Riemersma:1994hv} and found
agreement at the same level. The two codes agree equally well
for bottom quark production (Table~\ref{tab:xsec}).
As an independent cross-check,
Fig.~\ref{fig:nlo_hsmf} additionally overlays a \powheg variant with
a massified virtual
amplitude~\cite{Mitov:2006xs,Czakon:2007ej,Czakon:2007wk}: at $Q^2 = 12\GeV^2$, just above the
$4 m_c^2$ threshold, it visibly overshoots the full-mass result, as
expected from the breakdown of the $m_c \to 0$ expansion; at
$Q^2 = 200\GeV^2$ ($m_c^2/Q^2 \approx 0.01$) the three curves
coincide within Monte-Carlo statistics. 
These findings are consistent with observations on the asymptotic behavior of inclusive heavy-quark structure functions in the limit $Q^2/m^2 \gg 1$, as obtained from heavy-quark operator matrix elements and inclusive DIS structure functions in the massless-quark case; see Ref.~\cite{Ablinger:2025awb}.

\section{Charm cross sections at H1 and ZEUS}
%
\begin{figure}[b!]
  \centering
  \begin{subfigure}[t]{0.48\columnwidth}
    \includegraphics[width=\linewidth]{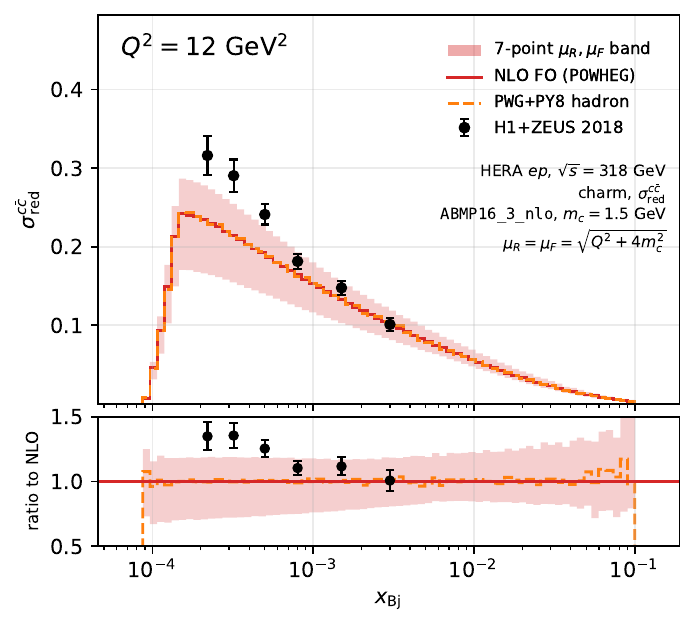}
    \caption{$Q^2 = 12\GeV^2$ (just above $4 m_c^2$).}
    \label{fig:sigred_charm_data_q2_12}
  \end{subfigure}\hfill
  \begin{subfigure}[t]{0.48\columnwidth}
    \includegraphics[width=\linewidth]{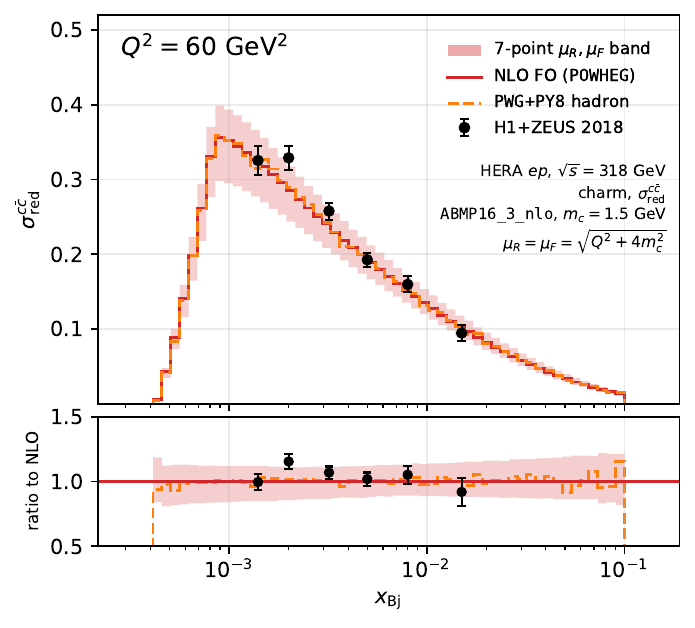}
    \caption{$Q^2 = 60\GeV^2$.}
    \label{fig:sigred_charm_data_q2_60}
  \end{subfigure}
  \caption{Charm reduced cross section $\sigma_{\rm red}^{c\bar c}$
    vs. $x_{\rm Bj}$ compared to the H1+ZEUS combined 2018
    measurement (filled circles): NLO fixed-order \powheg prediction
    (solid) with the 7-point $(\mu_R, \mu_F)$ scale-variation band
    (shaded), and the particle-level \powheg{}+\texttt{PYTHIA8}
    prediction (dashed). Lower panels show the same curves normalized
    to the NLO central, together with data/NLO at each data
    $x_{\rm Bj}$ (see text).}
  \label{fig:sigred_charm_data}
\end{figure}

Figure~\ref{fig:sigred_charm_data} compares our charm reduced
cross-section prediction to the H1+ZEUS combined 2018
measurement~\cite{H1:2018flt} at two representative $Q^2$. The NLO
fixed-order result is shown with its 7-point
$(\mu_R, \mu_F)$ scale-variation band, together with the
particle-level prediction
(\powheg{}+\texttt{PYTHIA8}~\cite{Bierlich:2022pnu}). The
hadronisation effect is negligible for this inclusive observable,
which is built from the scattered-electron kinematics (the electron
identified at Monte-Carlo truth level in the particle-level
analysis). At
$Q^2 = 60\GeV^2$ the data are reproduced within the scale uncertainty
across the full measured $x_{\rm Bj}$ range. At $Q^2 = 12\GeV^2$,
just above the $4 m_c^2$ threshold, the central prediction undershoots
the data at the lowest two $x_{\rm Bj}$ points by $\sim 30$--$40\%$,
exceeding the upper edge of the 7-point scale-variation band. The
scale uncertainty shrinks visibly from $Q^2 = 12$ to $60\GeV^2$, as
expected from perturbative convergence.

\section{Bottom-quark predictions}
%
\begin{figure}[b]
  \centering
  \begin{subfigure}[t]{0.48\columnwidth}
    \includegraphics[width=\linewidth]{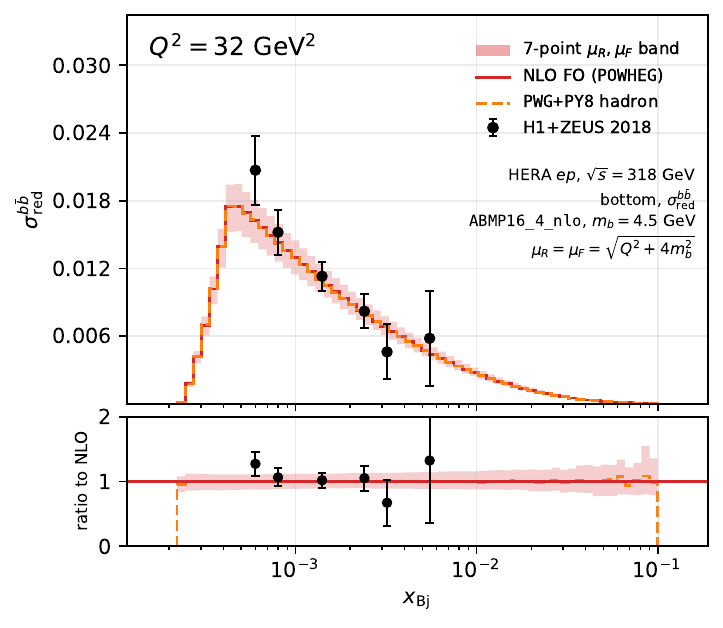}
    \caption{$Q^2 = 32\GeV^2$ (below $4 m_b^2 \approx 81\GeV^2$).}
    \label{fig:sigred_bottom_data_q2_32}
  \end{subfigure}\hfill
  \begin{subfigure}[t]{0.48\columnwidth}
    \includegraphics[width=\linewidth]{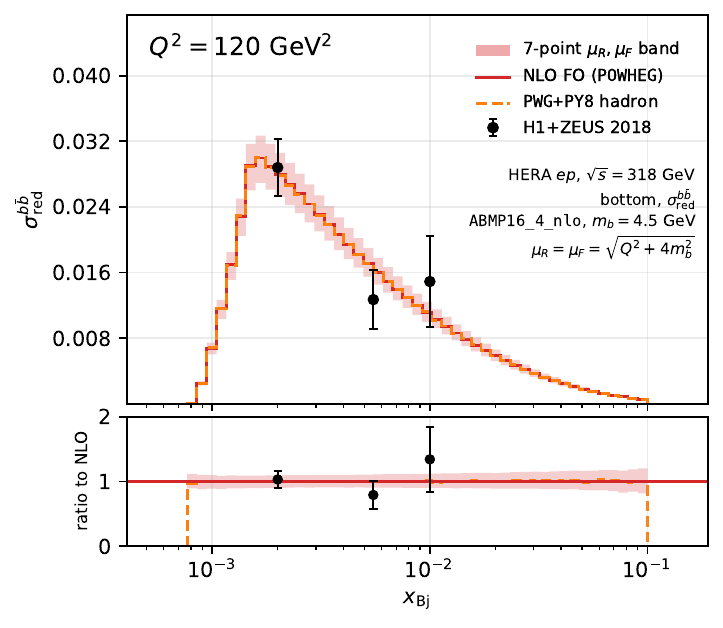}
    \caption{$Q^2 = 120\GeV^2$ (above $4 m_b^2$).}
    \label{fig:sigred_bottom_data_q2_120}
  \end{subfigure}
  \caption{Bottom reduced cross section $\sigma_{\rm red}^{b\bar b}$
    vs.\ $x_{\rm Bj}$ compared to the H1+ZEUS combined 2018
    measurement (filled circles): NLO fixed-order \powheg prediction
    (solid) and the particle-level \powheg{}+\texttt{PYTHIA8}
    prediction (dashed). Lower panels show the same curves normalized
    to the NLO central, together with data/NLO at each data
    $x_{\rm Bj}$ (see text).}
  \label{fig:sigred_bottom_data}
\end{figure}

Figure~\ref{fig:sigred_bottom_data} compares our 
$\sigma_{\rm red}^{b\bar b}$ prediction to the H1+ZEUS combined 2018
bottom quark measurement~\cite{H1:2018flt} at two representative $Q^2$ in
its 27-point grid: $Q^2 = 32\GeV^2$, sub-threshold ($4 m_b^2 \approx
81\GeV^2$, the regime where heavy-quark mass effects dominate), and
$Q^2 = 120\GeV^2$, above the threshold. The NLO fixed-order and the
particle-level (\powheg{}+\texttt{PYTHIA8}) predictions overlay
tightly: $\sigma_{\rm red}^{b\bar b}$ is left unchanged by the
shower at the bin level, and both curves describe the data within the
experimental uncertainty across both panels. The full set of
bottom quark differential distributions will be presented in a follow-up
paper.

\section{Hadron-level NLO predictions with \powheg}
%
\begin{figure}[t!]
  \centering
  \includegraphics[width=0.92\columnwidth]{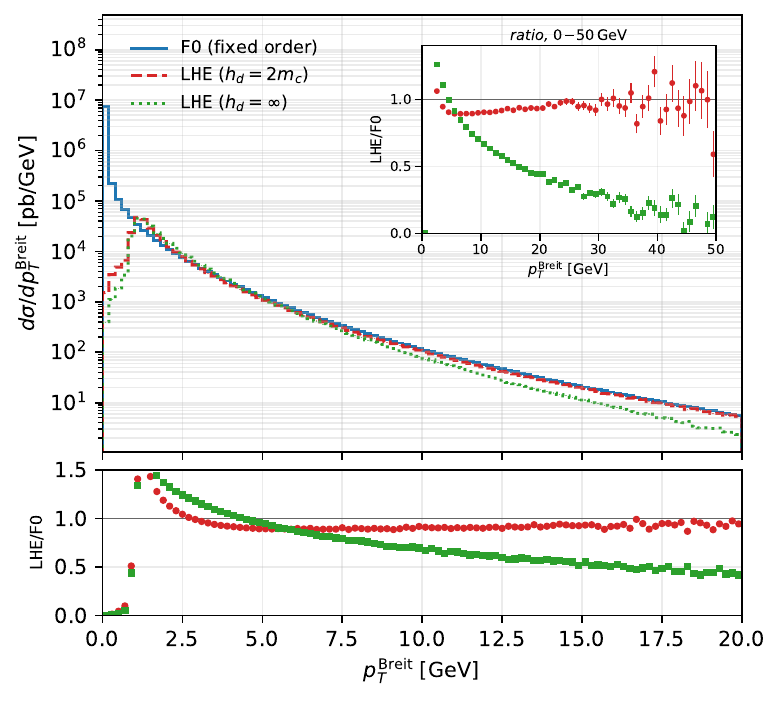}
  \caption{Extra-parton transverse momentum in the Breit frame, charm
    NLO with \powheg: fixed-order (solid), \powheg-matched LHE with
    damping factor $h_d = 2 m_c$ (dashed), and without damping
    (dotted). Lower panel: ratio between LHE and fixed-order (FO) prediction; inset: the same
    ratio over a wider $p_T^{\rm Breit}$ window (see text).}
  \label{fig:extra_pt}
\end{figure}

Figure~\ref{fig:extra_pt} shows the matching between the fixed-order
and the \powheg-matched (Les-Houches-event (LHE) stage) prediction for the extra-parton
transverse momentum. The observable is shown in the Breit frame
because the laboratory-frame transverse momentum picks up a $\sqrt{Q^2}$
floor from any $Q^2$ cut applied at generation, which the Breit
boost removes. \powheg routes the real-emission cross section
through the Sudakov-resummed $\tilde B$ piece at low $p_T$ and hands
the remainder back to the fixed-order remnant at high $p_T$; the
two pieces are interpolated by the damping factor
\begin{equation}
  F(p_T) \;=\; \frac{h_d^2}{h_d^2 + p_T^2}\, ,
  \label{eq:hdamp}
\end{equation}
where $p_T$ is the transverse momentum supplied internally by
\powheg in the underlying $2 \to 3$ real-emission kinematics
(distinct from the Breit-frame variant plotted in
Fig.~\ref{fig:extra_pt}), and $h_d$ is a scale that fixes the
transition~\cite{Alioli:2008tz}.
We adopt $h_d = 2 m_c$, the natural $c\bar c$ threshold of the
process. With this choice the LHE spectrum tracks the fixed-order
one across the hard region and converges to it in the wider-$p_T$
inset; without damping (the dotted curve) the LHE rate is
suppressed to $20$--$50\%$ of the fixed-order value throughout.

\section{Predictions for the EIC}
%
\begin{figure}[t!]
  \centering
  \begin{subfigure}[t]{0.48\columnwidth}
    \includegraphics[width=\linewidth]{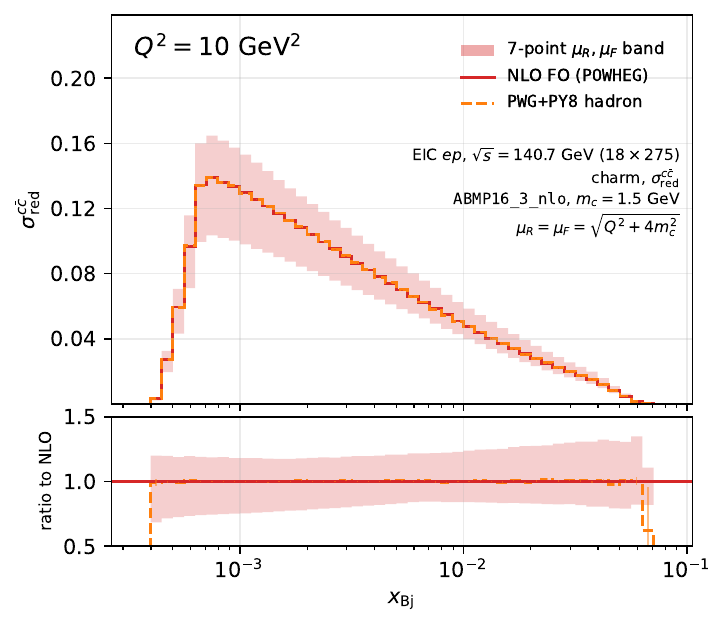}
    \caption{$Q^2 = 10\GeV^2$ (just above $4 m_c^2$).}
    \label{fig:eic_charm_q2_10}
  \end{subfigure}\hfill
  \begin{subfigure}[t]{0.48\columnwidth}
    \includegraphics[width=\linewidth]{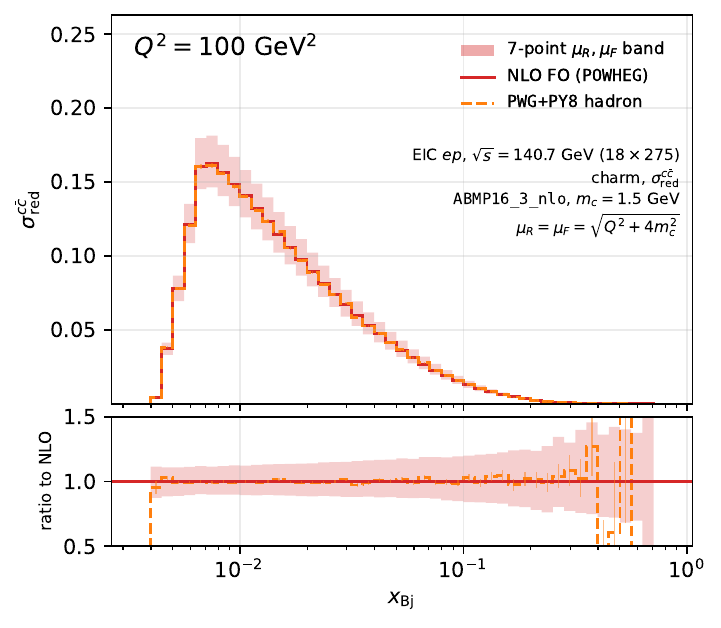}
    \caption{$Q^2 = 100\GeV^2$.}
    \label{fig:eic_charm_q2_100}
  \end{subfigure}
  \caption{Charm reduced cross section $\sigma_{\rm red}^{c\bar c}$
    vs.\ $x_{\rm Bj}$ at the EIC top energy
    ($\sqrt{s} \approx 141\GeV$): NLO fixed-order \powheg prediction
    (solid) and the particle-level \powheg{}+\texttt{PYTHIA8}
    prediction (dashed). Lower panels show the same curves normalized
    to the NLO central, together with data/NLO at each data
    $x_{\rm Bj}$ (see text).}
  \label{fig:eic_charm}
\end{figure}

Figure~\ref{fig:eic_charm} extends the charm prediction to the EIC
highest foreseen energy run, $E_e \times E_p = 18 \times 275\GeV$
($\sqrt{s} \approx 141\GeV$), at two representative $Q^2$ on a grid
with 5 bins per decade, following the Yellow Report
convention~\cite{AbdulKhalek:2021gbh}. The
fixed-order NLO and the particle-level (\powheg{}+\texttt{PYTHIA8})
predictions overlay tightly, confirming that the matching and shower
preserve the inclusive reduced cross section also in the EIC
kinematic envelope (the residual deviation at the sparse high-$x$
edges is statistical, not physical). The higher
luminosity available at the EIC~\cite{AbdulKhalek:2021gbh} will probe these predictions
with precision well beyond the HERA experimental uncertainty seen in
Fig.~\ref{fig:sigred_charm_data}.

\section{Conclusions}
In this Letter, we have presented the calculation and Monte Carlo implementation of heavy-quark pair-production in neutral-current DIS, reaching NLO+PS accuracy in QCD within the \powheg framework. This allows for a fully differential description of the process, including phase-space cuts on the final states, and is suitable for detailed studies of fragmentation, e.g. of $D$-mesons measured at HERA. It also complements existing \powheg implementations of heavy-quark production in charged-current DIS~\cite{Buonocore:2024pdv}.

As a proof of concept, the current implementation includes only the photon-gluon-fusion channel, which is numerically dominant in the kinematic range of HERA and the EIC, as evidenced by its good agreement with HERA data. 
For this channel, we have compared the NLO virtual corrections from \hvqdiscode~\cite{Harris:1997zq} with massified virtual corrections based on the light-quark virtual corrections. For larger values of $Q^2/m^2 \gg 1$, we find perfect agreement, which suggests a promising way to extend fully differential QCD corrections to heavy-quark DIS at NNLO in that kinematic regime.
The next steps include the completion of all channels, including quark-initiated ones, and dedicated simulations with parton-shower Monte Carlo generators (e.g. \texttt{PYTHIA8}~\cite{Bierlich:2022pnu} or \texttt{HERWIG}~\cite{Bewick:2023tfi}). 
These will provide valuable input for future EIC  Technical Design Report studies. The complete code will then be made available online.

Further possible directions include the use of different renormalization schemes for the heavy-quark mass. So far, we have used the on-shell scheme, with the usual caveats, especially for charm but also for bottom quarks. The conversion to the $\overline{\rm MS}$-scheme (running mass) is available~\cite{Alekhin:2010sv}. In view of the EIC capabilities regarding polarized beams, an extension to polarized heavy-quark pair production is also natural; the necessary NLO QCD corrections to polarized DIS are available~\cite{Hekhorn:2018ywm}.
Our formalism for the real-emission matrix elements in the full lepton-hadron scattering process readily accommodates such extensions.

Overall, these developments pave the way for precision studies of heavy-quark production in DIS at future electron-hadron collider facilities.

\section*{Acknowledgments}
We thank O.~Zenaiev for discussions and A.K. acknowledges the EIC
Theory Institute of the Brookhaven National Laboratory, where this
research was partially completed. 

We acknowledge support from the ERC
Advanced Grant 101095857 Conformal-EIC. 
C.D.P. is supported
by the U.S. Department of Energy under Contract No. DESC0012704.
A.K. is supported by the
University of Debrecen Program for Scientific Publication.

\bibliographystyle{apsrev4-1}
\bibliography{refs}

\end{document}